\documentclass[prl,twocolumn,showpacs,amsmath,amssymb,superscriptaddress,floatfix,nofootinbib]{revtex4-2}

\usepackage{graphicx}
\graphicspath{{figs/}}
\usepackage{dcolumn}
\usepackage{bm}
\usepackage{amssymb,mathtools}
\usepackage{booktabs}
\usepackage{siunitx}
\usepackage{multirow}
\usepackage{color}
\usepackage{microtype}
\usepackage[T1]{fontenc}
\usepackage{lmodern} 
\usepackage{enumitem}

\usepackage{physics}
\usepackage{upgreek}
\usepackage{amsmath}

\allowdisplaybreaks
\usepackage{placeins}

\usepackage[colorlinks,plainpages=false,linkcolor=blue,urlcolor=blue,citecolor=blue,pdfpagemode=UseNone,pdfstartview=FitBH]{hyperref}

\usepackage{sidecap}

\renewcommand{\tablename}{Table}
\makeatletter\renewcommand{\fnum@table}[1]{\tablename~\thetable.}\makeatother

\makeatletter

\newcommand{\LNO}{La$_3$Ni$_2$O$_{7}$}
\newcommand{\LNOd}[1]{La$_3$Ni$_2$O$_{#1}$}

\newcommand{\re}{\operatorname{Re}}

\definecolor{citecolor}{rgb}{0.0,0.60,0.32}

\begin{document}

\title{Dual instability of superconductivity from oxygen defects in La$_3$Ni$_2$O$_{7+\delta}$}

\author{Peiheng Jiang}
\thanks{These authors contributed equally to this work.}
\affiliation{School of Physics, MOE Key Laboratory for Nonequilibrium Synthesis and Modulation of Condensed Matter, Xi'an Jiaotong University, Xi'an 710049, China}

\author{Jie Li}
\thanks{These authors contributed equally to this work.}
\affiliation{National Laboratory of Solid State Microstructures and School of Physics, Nanjing University, Nanjing 210093, China}

\author{Yu-Han Cao}
\thanks{These authors contributed equally to this work.}
\affiliation{National Laboratory of Solid State Microstructures and School of Physics, Nanjing University, Nanjing 210093, China}

\author{Xiaodong Cao}
\affiliation{Suzhou Institute for Advanced Research, University of Science and Technology of China, Suzhou 215123, China}
\affiliation{School of Artificial Intelligence and Data Science, University of Science and Technology of China, Hefei 230026, China\looseness=-1}

\author{Zhicheng Zhong}
\affiliation{Suzhou Institute for Advanced Research, University of Science and Technology of China, Suzhou 215123, China}
\affiliation{School of Artificial Intelligence and Data Science, University of Science and Technology of China, Hefei 230026, China\looseness=-1}

\author{Yi Lu}
\email[]{yilu@nju.edu.cn}
\affiliation{National Laboratory of Solid State Microstructures and School of Physics, Nanjing University, Nanjing 210093, China}
\affiliation{Collaborative Innovation Center of Advanced Microstructures, Nanjing University, Nanjing 210093, China}

\author{Qiang-Hua Wang}
\email[]{qhwang@nju.edu.cn}
\affiliation{National Laboratory of Solid State Microstructures and School of Physics, Nanjing University, Nanjing 210093, China}
\affiliation{Collaborative Innovation Center of Advanced Microstructures, Nanjing University, Nanjing 210093, China}

\date{\today}

\begin{abstract}
We uncover a dual mechanism by which oxygen defects suppress superconductivity in the bilayer nickelate \LNOd{7+\delta} using density functional theory, dynamical mean-field theory, and functional renormalization group analysis. Apical vacancies and interbilayer interstitials emerge as the dominant low-energy defect species and are further stabilized by orthorhombic domain walls. These two defect classes drive the electronic structure in opposing directions. Vacancy-induced disorder generates local magnetic moments and promotes Anderson localization at moderate concentrations, whereas periodic interstitial ordering yields a coherent but weakly correlated metallic background that fails to support superconductivity. These findings highlight the decisive role of oxygen defects in shaping the superconducting and provide microscopic guidance for improving superconductivity through controlled defect engineering.
\end{abstract}

\maketitle

\emph{Introduction.---}
The discovery of superconductivity with $T_c \approx 80$ K in bulk \LNO{} under hydrostatic pressure~\cite{Sun2023,Hou2023,Zhang2024} has sparked intense interest in nickelate superconductors (see Refs.~\cite{Wang2024review,Wang2025review} for recent reviews). From the outset, oxygen stoichiometry emerged as a central factor, with early studies identifying oxygen vacancies~\cite{Dong2023} as a likely source of the low superconducting volume fractions in initial samples~\cite{Zhou2025investigations}. Despite steady improvements in synthesis, reliably stabilizing the pressurized superconducting phase remains challenging~\cite{Wang2024,Wang2024pressure,Li2025,Shi2025,Li2025ambient}. At ambient pressure, superconductivity has thus far been realized only in epitaxially strained thin films~\cite{Ko2024,Zhou2025}.

A natural approach to improving sample quality is to remove oxygen vacancies through post-growth annealing~\cite{Liu2025anneal,Zhang2025damage}. This procedure, however, has revealed a deeper puzzle. High-pressure oxygen annealing drives the material toward stoichiometric \LNO{} and improves ambient-pressure metallicity, yet it can simultaneously suppress superconductivity even under pressure~\cite{Shi2025,Huo2025}. This paradox indicates that restoring the average oxygen content is insufficient and that annealing can stabilize competing structural or electronic states that are unfavorable for pairing. Supporting this view, near-field microscopy reports stripe-like structural variations linked to local oxygen inhomogeneity~\cite{Zhou2025reveal}, and atomic-resolution imaging identifies interstitial oxygen in nonsuperconducting samples, implicating excess or redistributed oxygen as a key control parameter~\cite{Dong2025}. These observations point to oxygen as an active degree of freedom even in annealed samples, yet the microscopic connection between specific defect configurations and suppressed superconductivity remains unresolved.

Unraveling this connection requires determining how oxygen defects stabilize and how they reshape the correlated electronic landscape. In this Letter, we address this by combining density functional theory (DFT) to resolve defect energetics, dynamical mean-field theory (DMFT) to capture their impact on local correlations and coherence, and functional renormalization group (FRG) to assess the resulting superconducting tendencies. DFT identifies apical vacancies and interstitials as universal low-energy defects, further stabilized by orthorhombic domain walls at ambient pressure. DMFT shows that vacancies degrade electronic coherence and promote local magnetic moments, whose random distribution further drives Anderson localization. Conversely, interstitials arrange periodically to produce a coherent but weakly correlated metallic background. FRG reveals that this oxygen-rich metal fails to develop the robust $s_{\pm}$-wave pairing seen in stoichiometric \LNO. These results establish a dual mechanism for the suppression of superconductivity and indicate that the intrinsic high-$T_c$ phase is best realized by suppressing random vacancy disorder while strictly preventing interstitial accumulation.

\begin{figure}
  \includegraphics[width=\linewidth]{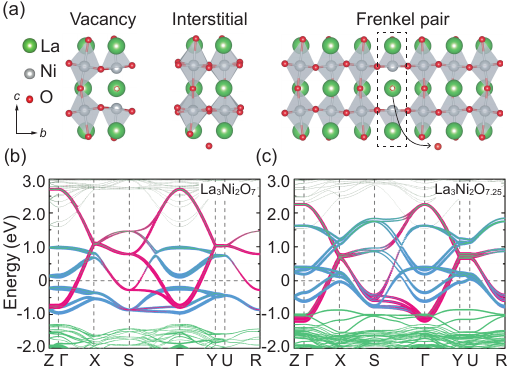}
  \caption{\label{fig:defects}
  (a) Schematic structures of oxygen defects in \LNO. The dashed box marks the boundary between two orthorhombic domains.
  (b, c) DFT+$U$ band structures of (b) pristine \LNO{} and (c) \LNOd{7.25}. Cyan, magenta, and green indicate Ni $d_{x^2-y^2}$, $d_{3z^2-r^2}$, and O~$p$ orbital character, respectively.}
\end{figure}

\emph{Oxygen vacancies and interstitials.---}
To determine the energetics of oxygen defects, we performed DFT(+$U$) calculations using the Vienna Ab initio Simulation Package (VASP)~\cite{Dudarev1998,VASP1996} with the Perdew-Burke-Ernzerhof (PBE) functional~\cite{PBE1996}. An effective $U=4.0$~eV was adopted, consistent with previous studies that reproduce the experimental crystal structure and Fermi surface~\cite{Sun2023,Yang2024}. Figure~\ref{fig:defects}(a) illustrates the defect types considered, namely vacancies, interstitials, and their combination into charge-neutral Frenkel pairs. Multiple candidate locations were examined for each defect type. Energetics identify the inner apical site as the most favorable position for a vacancy~\cite{Ni2025}, while interstitial oxygen preferentially resides between two nickel-oxygen bilayers. These configurations match the dominant defect species observed experimentally~\cite{Dong2023,Dong2025}. Further computational details are provided in the Supplemental Material (SM)~\cite{SM}.

The calculated formation energies show that \LNO{} readily accommodates both oxygen deficiency and excess. Vacancy formation is moderately costly (2.35~eV), comparable to high-valence transition-metal (e.g., Ni$^{3+}$ or Fe$^{4+}$) oxides that are prone to oxygen loss~\cite{Curnan2014}, implying that vacancies can persist in as-grown samples. By contrast, interstitial oxygen exhibits a large negative formation energy ($-1.22$~eV), indicating a strong thermodynamic tendency toward oxygen uptake.
As a result, the formation energy of a Frenkel pair is only $0.89$~eV, significantly lower than in benchmark oxide ion conductors~\cite{Huang2014}, pointing to a labile oxygen sublattice susceptible to intrinsic disorder.
Notably, the structural impact of these defects is markedly asymmetric. A vacancy contracts the $c$-axis by only 0.5\%, whereas an interstitial induces a substantial 3.5\% reduction~\cite{SM}.
The resulting elastic strain makes random vacancies relatively benign but drives interstitials to cluster and order, consistent with recent microscopy observations~\cite{Dong2025}.

These defect energetics are further reshaped by extended structural motifs, most notably domain walls. At ambient pressure, orthorhombic \LNO{} hosts low-energy domain walls formed by reversals of the octahedral rotation pattern along the $b$ axis [Fig.~\ref{fig:defects}(a)], stabilized at only $3.5$meV/\AA$^2$~\cite{SM}. Our calculations identify these boundaries as effective defect traps. The formation energy of a vacancy drops by $0.37$~eV when located on the wall, whereas that of an interstitial decreases by $0.23$~eV when placed adjacent to the wall. This cumulative stabilization reduces the local Frenkel pair energy to just $0.31$~eV. This low energy scale indicates that domain walls strongly promote the formation and pinning of nearby oxygen defects during synthesis or annealing, embedding intrinsic inhomogeneity even when the global composition is nominally stoichiometric. Such defect accumulation provides a plausible microscopic origin for the stripe-like conductivity modulations observed in \LNOd{7+\delta}~\cite{Zhou2025reveal}.

\begin{table}[t]
\caption{\label{tab:dftdmft}
Crystal-field splittings and orbital-resolved occupations for \LNOd{7+\delta}$ (\delta = 0, \pm0.5$) at 0 (top) and 20~GPa (bottom). DMFT results were obtained at $T=116$~K. Ill-defined entries are shown as slashes.}
\begin{ruledtabular}
\begin{tabular}{c c @{\hspace{12pt}}
                S[table-format=1.2]
                S[table-format=1.2]
                S[table-format=1.2]
                @{\hspace{12pt}}
                S[table-format=1.2]
                S[table-format=1.2]
                S[table-format=1.2]
                S[table-format=1.2]}
\multicolumn{2}{c}{} & \multicolumn{3}{c}{DFT} & \multicolumn{4}{c}{DMFT} \\
$\delta$ & site &
{$\Delta$} &
{$n_x$} &
{$n_z$} &
{$n_x$} &
{$n_z$} &
{$Z_x$} &
{$Z_z$} \\
\hline
\multicolumn{9}{l}{0~GPa} \\[4pt]
0 &  & 0.25 & 0.67 & 0.83 & 0.67 & 0.83 & 0.42 & 0.27 \\
-0.5 & pyr. & 1.55 & 0.49 & 1.98 & 0.96 & 1.03 & \multicolumn{1}{c}{/} & 0.23 \\
-0.5 & oct. & 0.01 & 0.63 & 0.90 & 1.01 & 1.00 & \multicolumn{1}{c}{/} & \multicolumn{1}{c}{/} \\
0.5 &  & -0.09 & 0.63 & 0.37 & 0.61 & 0.39 & 0.54 & 0.66 \\
\hline
\multicolumn{9}{l}{20~GPa} \\[4pt]
0 &  & 0.33 & 0.63 & 0.87 & 0.66 & 0.84 & 0.55 & 0.36 \\
-0.5 & pyr. & 1.79 & 0.52 & 1.97 & 0.49 & 1.58 & 0.29 & 0.39 \\
-0.5 & oct. & 0.04 & 0.60 & 0.91 & 0.96 & 0.97 & 0.15 & 0.19 \\
0.5 &  & -0.10 & 0.63 & 0.37 & 0.61 & 0.39 & 0.60 & 0.70 \\
\end{tabular}
\end{ruledtabular}
\end{table}

\begin{figure*}[t]
  \includegraphics[width=\linewidth]{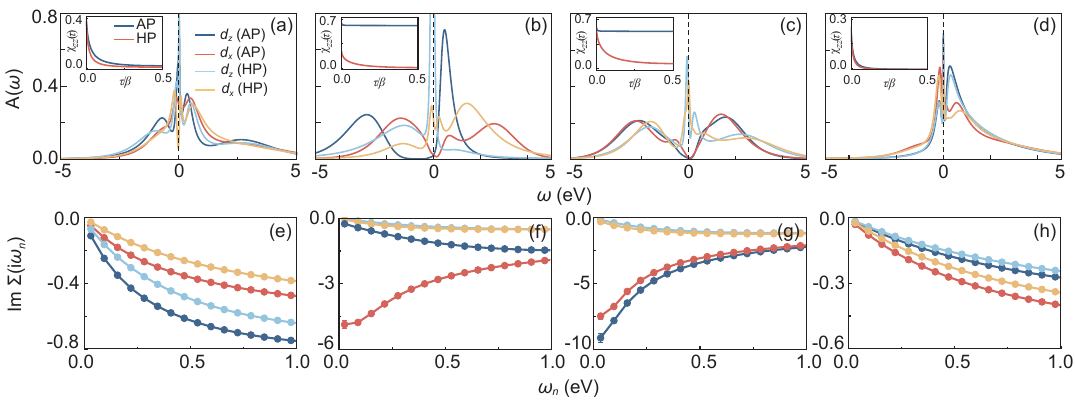}
  \caption{\label{fig:dmft}
  DMFT spectral functions (a-d) and self-energies (e-h) for \LNOd{7} [(a),(e)], \LNOd{6.5} [(b),(f) pyramidal sites; (c),(g) octahedral sites], and \LNOd{7.5} [(d),(h)] under ambient pressure (AP, 0 GPa) and high pressure (HP, 20 GPa). Insets in panels (a-d) show the imaginary-time spin-spin correlation functions $\chi_{zz}(\tau) = \langle S_z(\tau) S_z(0) \rangle$, where $S_z$ is the total $z$ spin component of the $d_x$ and $d_z$ orbitals.}
\end{figure*}

\emph{Single-particle electronic structure with defects.---}
The oxygen vacancies and interstitials reshape the local electronic states of \LNOd{7+\delta} through distinct lattice responses.
Figures~\ref{fig:defects}(b,c) compare the DFT+$U$ band structure of pristine \LNO{} with those of an interstitial-rich configuration \LNOd{7.25} in which one interbilayer region is fully occupied by interstitial oxygen.
Interstitials shorten the surrounding apical Ni-O bonds and strongly increase interorbital hybridization, pushing the $d_{3z^2-r^2}$ (hereafter $d_z$) bands upward relative to $d_{x^2-y^2}$ ($d_x$) and enhancing their dispersion.
Conversely, removing an inner-apical oxygen converts adjacent NiO$_6$ octahedra into NiO$_5$ square pyramids [Fig.~\ref{fig:defects}(a)], lowers the local $d_z$ level and suppresses intrabilayer hybridization.
The characteristic $d_z$ bonding-antibonding splitting then collapses, and the associated $d_z$ states sink to the bottom of the $e_g$ manifold, well below the Fermi level~\cite{SM}.

To quantify these changes, we constructed maximally localized Wannier functions~\cite{wien2k,Mostofi2014} for $d_x$ and $d_z$ and extracted crystal-field splittings and orbital occupations for the limiting oxygen-deficient (\LNOd{6.5}) and oxygen-rich (\LNOd{7.5}) structures (Table~\ref{tab:dftdmft}).
In pristine \LNO, the splitting $\Delta=\epsilon_x-\epsilon_z$ is modest, producing a moderate polarization toward $d_z$ with $n_z = 0.83$ close to half filling.
In the vacancy-rich \LNOd{6.5}, the pyramidal sites develop a large positive splitting ($\Delta = 1.55$ eV) that nearly saturates the $d_z$ orbital ($n_z = 1.98$), whereas the octahedral sites remain similar to the stoichiometric limit with a weaker anisotropy.
The imbalance in local hybridizations results in a pronounced charge disproportionation, with about $2.5$ $e_g$ electrons on the pyramidal sites and $1.5$ on the octahedral sites.
Oxygen-rich \LNOd{7.5}, on the other hand, inverts the ordering of the $e_g$ levels ($\Delta=-0.09$~eV), driving $n_z$ down to $0.37$ as the total $e_g$ filling drops to 1.0 per Ni.
These crystal-field shifts, hybridization changes, and charge redistributions persist under high pressure without altering the qualitative hierarchy for all the three compositions.

\emph{Correlated electronic structure.---}
Building on the DFT analysis, we now examine how oxygen defects reshape the correlated electronic states of \LNOd{7+\delta}.
DMFT calculations were performed using the TRIQS library~\cite{TRIQS2015,CTHYB2016,DFTTools2016,maxent} with a rotationally invariant Kanamori interaction ($U=3.4$ eV, $J_H=0.6$ eV~\cite{Christiansson2023}) at $T=116$~K (inverse temperature $\beta=100$).
For pristine \LNO{} [Fig.~\ref{fig:dmft}(a)], the spectral functions remain qualitatively robust against pressure, showing typical characteristics of a Hund's metal~\cite{Georges2013review}, with narrow quasiparticle peaks at the Fermi level and broad Hubbard bands at higher energies~\cite{Christiansson2023,Shilenko2023,Lechermann2023,Cao2024,Ouyang2024hund,Wang2024electronic}.
The imaginary parts of the self-energies [Fig.~\ref{fig:dmft}(e)] are reduced under high pressure, indicating bandwidth-driven weakening of correlations. They remain larger for $d_z$ than for $d_x$, showing pronounced orbital differentiation. The quasiparticle weights $Z=[1-\partial\re\Sigma(i\omega)/\partial \omega|_{\omega\rightarrow 0}]^{-1}$ are extracted as $Z_x = 0.42$ and $Z_z = 0.27$ at ambient pressure. These correspond to mass enhancements of about 2 and 4, respectively, in good agreement with angle-resolved photoemission spectroscopy measurements~\cite{Yang2024}.

For the vacancy-rich \LNOd{6.5}, correlations drastically reshape the electronic landscape. As shown in Table~\ref{tab:dftdmft}, local Coulomb interactions largely suppress the DFT-predicted charge disproportionation, restoring a filling of $n \approx 2$ on both site types and eliminating the strong orbital polarization at the pyramids.
At ambient pressure, the spectral function [Fig.~\ref{fig:dmft}(b)] reveals that the $d_z$ orbital on the pyramidal sites remains metallic but with vanishing quasiparticle spectral weight, whereas the $d_x$ orbital is pushed to the verge of a Mott transition, consistent with its nearly divergent self-energy [Fig.~\ref{fig:dmft}(f)].
In contrast, the octahedral sites become fully Mott insulating, displaying gapped spectra and divergent self-energies [Figs.~\ref{fig:dmft}(c,g)].
The spin correlation function $\chi_{zz}(\tau) \equiv \langle S_z(\tau)S_z(0)\rangle$ reveals robust frozen moments on both sublattices. The equal-time value $\chi_{zz}(0) \approx 2/3$ corresponds to an instantaneous effective moment $S_{\mathrm{eff}} \approx 1$, while the negligible decay at $\tau=\beta/2$ signifies that these moments are essentially unscreened~\cite{Werner2008spinfreezing}.
The formation of these robust local moments is a direct consequence of the disrupted intrabilayer hybridization at the pyramidal sites combined with the effective doping towards half-filling. Under high pressure, a uniform vacancy lattice maintains sufficient coherence to screen these moments and recover a metallic state [Figs.~\ref{fig:dmft}(b,c) and (f,g)]. However, as our DFT analysis implies, the vacancies are randomly distributed in realistic samples. As detailed in the SM~\cite{SM}, modeling these local moments as random scatterers reveals that the resulting disorder drives the system into an Anderson localized state at moderate concentrations. This mechanism naturally accounts for the experimentally observed magnetic insulating behavior in oxygen-deficient samples, even under pressure~\cite{Gao2024}, and aligns with the Anderson-localization characteristics found in transport measurements~\cite{Poltavets2006}.

In oxygen-rich \LNOd{7.5}, the reduced $e_g$ filling weakens electronic correlations relative to \LNO{}. The Hubbard bands are strongly suppressed in the spectral function [Fig.~\ref{fig:dmft}(d)], and the imaginary parts of the self-energies are much smaller [Fig.~\ref{fig:dmft}(h)].
Orbital differentiation is effectively eliminated, yielding large and nearly equal quasiparticle weights (Table~\ref{tab:dftdmft}).
The increase in $Z$ translates into a higher renormalized Fermi velocity $v_F^\ast \simeq Z v_F$ and a reduced quasiparticle scattering rate $\tau^{-1}\approx -2Z\mathrm{Im}\Sigma(0^+)$, signaling a highly coherent metal.
This coherence is further reflected in the fully screened local spin response. The static impurity susceptibility $\chi_{\mathrm{imp}}=\int_0^\beta d\tau\,\chi_{zz}(\tau)$ drops to $1.3$ and $1.0$ eV$^{-1}$ at 0 and 20 GPa, respectively, much smaller than in pristine \LNOd{7} under the same conditions ($5.2$ and $3.0$ eV$^{-1}$).
These results suggest that oxygen intercalation shifts the system toward the weak-coupling limit, placing \LNOd{7.5} in a parameter regime where low-energy instabilities arise, if at all, from residual Fermi-surface interactions rather than from strong local correlations.

\emph{Oxygen-tuned superconductivity.---}
Motivated by the weak correlation revealed by DMFT for \LNOd{7.5}, we employ singular-mode FRG (SM-FRG)~\cite{Wang2012,Xiang2012,Wang_PRB_2013} to analyze its low-energy instabilities and superconducting tendencies relative to pristine \LNO.
In SM-FRG, the one-particle-irreducible four-point vertex $\Gamma_\Lambda$ is projected onto scattering matrices between fermion bilinears in the superconducting (SC), spin-density-wave (SDW), and charge-density-wave (CDW) channels, whose flow is then tracked as a function of a running infrared cutoff $\Lambda$. An instability is signaled when the most negative singular value $S_\Lambda$ in one channel diverges at a critical scale $\Lambda_c$, which serves as an estimate of the associated transition temperature. This approach provides an unbiased treatment of competing ordering channels in the weak-to-moderate interaction regime~\cite{Metzner_RMP_2012,Berges_PR_2002,Dupuis_PR_2021,Kopietz__2010} and has previously captured the $s_\pm$ superconducting state in bilayer and trilayer nickelates and its pressure evolution~\cite{Yang2023,Yangqg_prb_2024_4310,Jiang2025,Cao_SCPMA_2025}. Additional details are provided in the SM~\cite{SM}.

\begin{figure}
  \includegraphics[width=\linewidth]{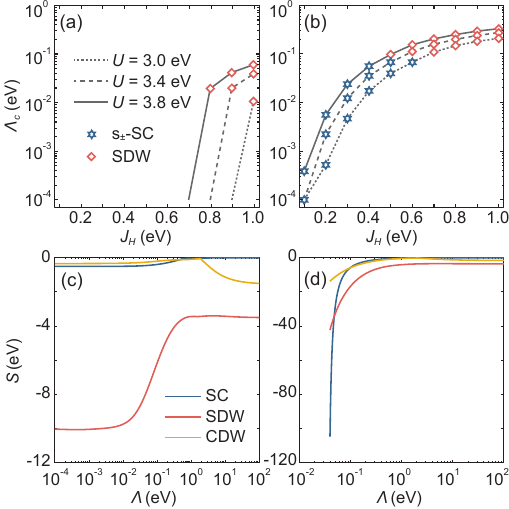}
  \caption{\label{fig:frg}
  Critical scales $\Lambda_c$ as a function of $J_H$ for different $U$ values in (a) \LNOd{7.5} and (b) \LNOd{7}. Panels (c) and (d) show representative flows of the singular values $S$ in the SC, SDW, and CDW channels for $(U, J_H) = (3.0, 0.5)$~eV in \LNOd{7.5} and \LNOd{7}, respectively.}
\end{figure}

The calculations use the same Kanamori interaction form as in DMFT, with $U$ and $J_H$ varied to map the leading electronic instabilities. The resulting phase diagram for \LNOd{7.5} is shown in Fig.~\ref{fig:frg}(a), with that of stoichiometric \LNO{} presented for comparison in Fig.~\ref{fig:frg}(b). For \LNOd{7.5}, no divergence is found in any channel down to the lowest cutoff $\Lambda = 10^{-4}$ eV, implying that any instability, if present, would occur only below $T_c \sim 1$ K. A noticeable SDW tendency appears only at larger $J_H$. Pristine \LNO{}, by contrast, hosts a robust $s_\pm$-wave superconducting phase at small and moderate $J_H$, and similarly develops an SDW instability at larger $J_H$, but at substantially higher energy scales. This comparison makes clear that oxygen-rich \LNOd{7.5} shows a pronounced suppression of both magnetic and superconducting tendencies relative to pristine \LNO{}.

To further clarify the origin of the contrasting phase diagrams, we show representative SM-FRG flows for \LNOd{7.5} and \LNO{} in Figs.~\ref{fig:frg}(c) and \ref{fig:frg}(d), using $U = 3.0$ eV and $J_H = 0.5$ eV. Both systems behave similarly at large cutoff $\Lambda$. The bare Coulomb repulsion suppresses the CDW channel, seen in the decreasing $S_{\mathrm{CDW}}$, and it prevents the development of an attractive SC channel, leaving the SDW channel dominating the initial flow. Around $\Lambda \sim 1$ eV, spin fluctuations strengthen the SDW channel and, through interchannel coupling, also enhance the CDW and SC channels. The crucial difference emerges at lower energy. In \LNOd{7.5}, all three channels eventually saturate and no divergence appears down to $10^{-4}$ eV, which leads to the absence of an instability in Fig.~\ref{fig:frg}(a). In \LNO{}, however, the SDW fluctuations remain much stronger and significantly reinforce the SC channel through the Kohn-Luttinger mechanism~\cite{Kohn-Luttinger_1965}, ultimately driving a Cooper instability. Together, these flow behaviors show that excess oxygen suppresses the superconducting instability in \LNOd{7.5} by weakening the SDW channel and the associated spin fluctuations that mediate pairing. Notably, the residual SDW tendency in \LNOd{7.5} favors interlayer \emph{ferromagnetic} correlations, in stark contrast to the interlayer antiferromagnetic pattern of pristine \LNO{}, providing a clear magnetic signature of oxygen-rich regions.

\emph{Summary and discussion.---}
Our analysis establishes that high-$T_c$ superconductivity in \LNOd{7+\delta} is confined to a narrow stability window bounded by two defect-driven regimes. On one side, thermodynamically accessible oxygen vacancies degrade coherence and introduce magnetic disorder, which drives Anderson localization at moderate concentrations. On the other, the accumulation of interstitials stabilizes a coherent but weakly correlated Fermi liquid where the correlations essential for pairing are diluted. This microscopic dichotomy resolves the paradox of oxygen-annealed samples, where aggressive annealing successfully removes fatal vacancies but can inadvertently populate interstitial sites, driving the system into a nonsuperconducting metallic state.

Practical optimization therefore demands breaking this trade-off. Beyond precise control of oxygen partial pressure during synthesis, our structural analysis suggests two specific routes for defect engineering. First, since interstitials induce a substantial $c$-axis contraction, applying strain to oppose this distortion could destabilize interstitial formation without hindering vacancy removal. Second, given that domain walls significantly lower defect formation energies, detwinning protocols~\cite{Liang2012detwin} could eliminate these energetic traps, thereby widening the window for stoichiometric superconductivity.

\begin{acknowledgments}
\emph{Acknowledgments.---}We thank Jian-Xin Li, Meng Wang, Motoharu Kitatani, and Qing Li for insightful discussions.
This work was supported by the National Key R\&D Program of China (Nos. 2022YFA1403000, 2024YFA1408100) and the National Natural Science Foundation of China (Nos. 12274207, 12374147, 12574069, 92365203). P. J. also acknowledges support from the Natural Science Basic Research Program of Shaanxi Province (No. 2025JC-YBQN-001), the Young Talent Support Project of Xi'an Jiaotong University (WL6J022) and the HPC Platform at Xi'an Jiaotong University.
\end{acknowledgments}

\emph{Data availability.---} Data supporting this manuscript are available in the Supplemental Material~\cite{SM}. Additional data are available from the corresponding authors upon reasonable request.


\end{document}